\documentclass[jap,amssymb,preprint,superscriptaddress,showpacs]{revtex4}   
\usepackage{graphicx}

\begin{document}

\title{Magnetism of NiMn$_{2}$O$_{4}$-Fe$_{3}$O$_{4}$ Spinel Interfaces}

\author{B.B. Nelson-Cheeseman}
\email{bbnelsonchee@berkeley.edu} \affiliation{Department of
Materials Science and Engineering, University of California,
Berkeley, CA 94720}
\author{R.V. Chopdekar}
\affiliation{School of Applied and Engineering Physics, Cornell
University, Ithaca, NY 14853} \affiliation{Department of Materials
Science and Engineering, University of California, Berkeley, CA
94720}
\author{J.S. Bettinger}
\affiliation{Department of Materials Science and Engineering,
University of California, Berkeley, CA 94720}
\author{E. Arenholz}
\affiliation{Advanced Light Source, Lawrence Berkeley National
Laboratory, Berkeley, CA 94720}
\author{Y. Suzuki}
\affiliation{Department of Materials Science and Engineering,
University of California, Berkeley, CA 94720}

\date{\today}

\begin{abstract}
We investigate the magnetic properties of the isostructural
spinel-spinel interface of NiMn$_{2}$O$_{4}$(NMO)-Fe$_{3}$O$_{4}$.
Although the magnetic transition temperature of the NMO film is
preserved, both bulk and interface sensitive measurements
demonstrate that the interface exhibits strong interfacial magnetic
coupling up to room temperature. While NMO thin films have a
ferrimagnetic transition temperature of 60K, both NiFe$_{2}$O$_{4}$
and MnFe$_{2}$O$_{4}$ are ferrimagnetic at room temperature. Our
experimental results suggest that these magnetic properties arise
from a thin interdiffused region of (Fe,Mn,Ni)$_{3}$O$_{4}$ at the
interface leading to Mn and Ni magnetic properties similar to
MnFe$_{2}$O$_{4}$ and NiFe$_{2}$O$_{4}$.
\end{abstract}

\maketitle

The oxide spinel Fe$_{3}$O$_{4}$ is an ideal candidate for highly
spin polarized electrode material to be used in spintronic
applications. It has been theoretically predicted to be half
metallic, and is highly attractive for applications due to its high
Curie temperature (T$_C$) of ~850K\cite{Zhang91}. Experimental
studies of Fe$_{3}$O$_{4}$ in spintronic heterostructures, however,
have exhibited much lower junction magnetoresistance (JMR) values
than expected from a half-metallic electrode material. Among the
highest JMR values of Fe$_{3}$O$_{4}$-based heterostructures are
observed in layered systems with epitaxially grown isostructural
spinel barrier layers. Oxide spinels such as CoCr$_{2}$O$_{4}$,
MgTi$_{2}$O$_{4}$, FeGa$_{2}$O$_{4}$, and MnCr$_{2}$O$_{4}$ have
been used as barrier layers in magnetic tunnel junctions with spinel
Fe$_{3}$O$_{4}$ and half metallic perovskite
electrodes\cite{Hu02,Alldredge06}, while CoFe$_{2}$O$_{4}$ has been
used with Fe$_{3}$O$_{4}$ in spin-filter
junctions\cite{Chapline06,Ramos07}. Recently, NiMn$_{2}$O$_{4}$
(NMO) has also been identified as an effective spin filter barrier
material in Fe$_{3}$O$_{4}$-based magnetic junctions with perovskite
counter electrodes\cite{N-C07}. Whereas perovskite and spinel layers
have been shown to be magnetically uncoupled in these
structures\cite{N-C07}, the magnetism near the isostructural spinel
interfaces is a subject of interest. A more detailed investigation
of the interfacial magnetic interactions between spinel structure
materials is necessary in order to understand transport and magnetic
interaction results attributed to these multilayers, as well as to
optimize the use of these heterostructures for spintronic
applications.

In this paper, we observe magnetic properties in NiMn$_{2}$O$_{4}$
thin film bilayers with Fe$_{3}$O$_{4}$ not observed in either film
alone. Although the NMO magnetic transition at 60K is preserved,
interfacial element-specific magnetism measurements of
NMO/Fe$_{3}$O$_{4}$ bilayers show strong interfacial coupling of the
Fe, Mn and Ni moments. We suggest these magnetic results can be
explained by a thin interdiffused layer at the interface.

Thin film heterostructures of NMO and Fe$_{3}$O$_{4}$ were grown by
pulsed laser deposition on (110)-oriented single crystal SrTiO$_{3}$
(STO) substrates. The NMO was grown at 600$^\circ$C in 10mTorr of
99\%N$_{2}$/1\%O$_{2}$, while the Fe$_{3}$O$_{4}$ was grown at
400$^\circ$C in vacuum. The NMO film was grown first to minimize
oxidation of the Fe$_{3}$O$_{4}$ film during deposition. The films
grow epitaxially on the STO substrates as confirmed by X-ray
diffraction and Rutherford Backscattering mesasurements. The single
NMO thin films have a T$_C$ of 60K\cite{unpubNMO}. The bulk
magnetism of the samples was probed by a superconducting quantum
interference device (SQUID) magnetometer. The element specific
magnetic properties of the interfacial Ni, Mn and Fe were
investigated by X-ray magnetic circular dichroism (XMCD) (BL4.0.2
and BL6.3.1) in total electron yield at the Advanced Light Source.
Due to the surface sensitive nature of this technique and in order
to be interface specific, all samples had a 5nm Fe$_{3}$O$_{4}$ top
layer. Additionally, because the NMO films have a low saturation
magnetization (0.8$\mu_{B}$) compared to Fe$_{3}$O$_{4}$ films
(4.1$\mu_{B}$), two different thicknesses of NMO film in the bilayer
(40nm and 5nm) were utilized to elucidate any effect of the bulk NMO
film on the interface. Lastly, because these measurements are
relevant to spin polarized heterostructures, where the bottom spinel
layer is usually grown on a perovskite counter electrode, such a
heterostructure was also investigated. Therefore, the
NMO/Fe$_{3}$O$_{4}$ interface was investigated in the following
three samples: a 'thick bilayer' of
STO//NMO(40nm)/Fe$_{3}$O$_{4}$(5nm), a 'thin bilayer' of
STO//NMO(5nm)/Fe$_{3}$O$_{4}$(5nm), and a 'trilayer' of
STO//La$_.7$Sr$_.3$MnO$_3$(LSMO)(40nm)/NMO(5nm)/Fe$_{3}$O$_{4}$
(5nm). All magnetic measurements were performed along the [100]
in-plane direction.

Moment versus temperature measurements taken at 10 Oe of the
NMO/Fe$_3$O$_4$ bilayers are shown in Fig 1(a). The thick bilayer
sample shows a Brillouin shape for the NMO T$_{C}$ of 60K; however,
after reaching a minimum at 60K, the moment begins to rise with
increasing temperature (Fig 1a), uncharacteristic of the magnetic
behavior observed in either individual film. This behavior is
largely absent in the thin bilayer sample, although a slight
discontinuity may be seen at ~50K (Fig 1(b)). Such results prompted
more detailed investigation of the magnetic interactions at the
interface.
\begin{figure}
\center{\includegraphics[width=6.5 cm]{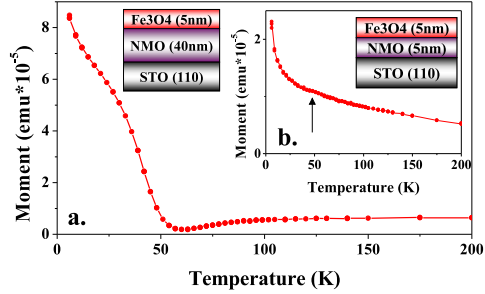}} \caption{Moment
as a function of temperature for (a) thick (40nm) NMO bilayer, and
(b) thin (5nm) NMO bilayer.}
\end{figure}

XMCD spectra and hysteresis loops were taken of the
NMO/Fe$_{3}$O$_{4}$ interface in all heterostructures at various
temperatures. The thin NMO bilayer XAS and XMCD results are
displayed in Fig 2. The NMO/Fe$_{3}$O$_{4}$ interface exhibits
virtually identical Fe, Mn and Ni XAS and XMCD spectra for all
temperatures between 30K-300K, as seen in Fig 2. The thick NMO
bilayer and trilayer samples also demonstrate this behavior. In
addition, for a given temperature, the Fe, Mn and Ni XMCD hysteresis
loops are identical to one another. Nevertheless, the \textit{shape}
of the hysteresis loops changes distinctly as a function of
temperature, showing a dramatic increase in coercive field for all
three elements below the NMO T$_C$. Similar results are seen in the
temperature dependent hysteresis loops of the trilayer sample. As
shown in Fig 3, the coercive fields of the Fe are the same at 80K
and 55K, but increase at 30K and 15K. Furthermore, at 55K, the
hysteresis loop shows a slight decrease in remanent dichroism, which
is consistent with the minimum moment in the SQUID data. As the
normalized Fe, Mn and Ni hysteresis loops are identical for each
given temperature, this hysteresis loop behavior is seen in the Mn
and Ni as a function of temperature also, but has been omitted for
clarity in Fig 3.
\begin{figure}
\center{\includegraphics[width=8 cm]{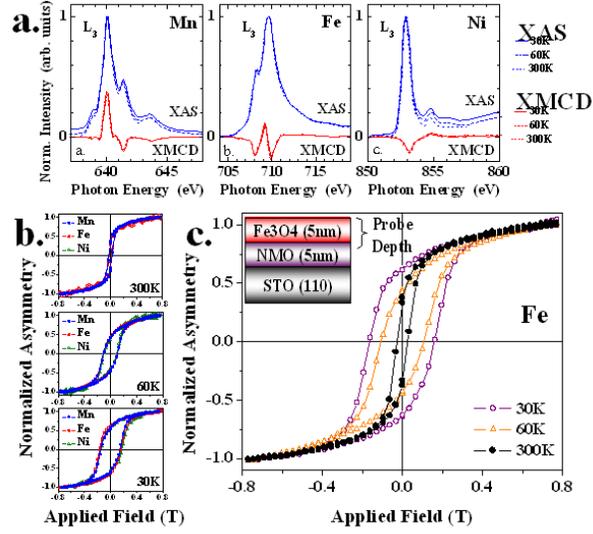}}
\caption{Element specific magnetism of Fe$_3$O$_4$/NMO interface in
thin NMO/Fe$_3$O$_4$ bilayer. (a) XAS and XMCD spectra for Mn, Fe,
and Ni as a function of temperature, (b) Mn, Fe and Ni normalized
XMCD hystersis loops at each temperature, (c) Fe normalized XMCD
hysteresis loops a function of temperature.}
\end{figure}

One can now discuss the apparent source of the bulk moment
measurements by utilizing the element and interface specific XMCD
information. First, it appears that there is significant magnetic
coupling at the interface as evidenced by identical Fe, Mn and Ni
hysteresis loops. However, although they are identical for a given
temperature, the magnetic nature of the hysteresis loops becomes
increasingly harder as the temperature is decreased through 60K.
This evolution suggests that the species at the interface are
coupled to the magnetically soft Fe$_3$O$_4$ at temperatures above
the NMO T$_C$, but, as the NMO layer becomes ferrimagnetic, the
species couple to the magnetically hard NMO, as well. The decrease
in remanent asymmetry observed in the trilayer around 60K could be
due to a magnetic frustration of the interfacial species as the NMO
layer becomes ferrimagnetic. This can all be related to the
\textit{increase} in bulk moment observed above 60K in Fig 1a in the
following way: just above the T$_{C}$ of the NMO film, the bulk
hysteresis loop exhibits greater squareness than that at lower
temperatures, which results in an effective \textit{increase} in
moment at small fields as the temperature is increased.
\begin{figure}
\center{\includegraphics[width=6.5 cm]{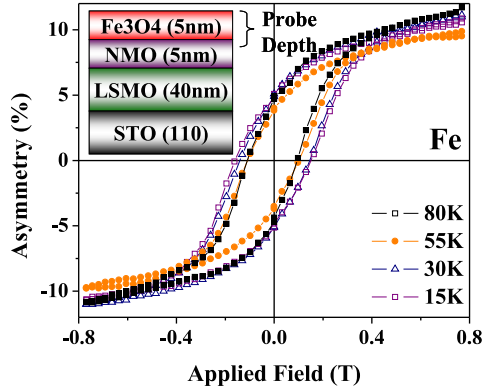}}
\caption{Fe XMCD hysteresis loops as a function of temperature,
probing the top ~5nm of the trilayer sample.}
\end{figure}

The magnetic transition of the NMO film in the presence of this
strong magnetic coupling at the interface is also of interest. It is
apparent from the change in hysteresis loop shape and increase in
coercive field that even the thin NMO layer undergoes a magnetic
transition around 60K. Any depressed onset of the coupling to the
NMO layer could be due to the relatively low magnetization of the
NMO compared to the Fe$_3$O$_4$.

Now that we have discussed how the interfacial species respond to
the bulk of the NMO and Fe$_3$O$_4$ thin films, let us focus on how
the magnetic species at the interface can give rise to room
temperature Mn and Ni magnetic circular dichroism. Two possible
explanations are: (1) a thin interfacial layer of the NMO thin film
is magnetized far above the NMO T$_C$ by the close proximity to the
Fe$_3$O$_4$ layer, or (2) the presence of a mixed
(Fe,Mn,Ni)$_3$O$_4$ spinel at the interface that is ferrimagnetic at
room temperature. Such a solid solution at the interface is
reasonable as the cations of the spinel structure occupy only a
small fraction of the available octahedral and tetrahedral sites of
the oxygen face-centered-cubic sub-lattice, leaving ample
opportunity for cation diffusion throughout the structure.

The XAS and XMCD data supports the presence of Mn and Ni in
MnFe$_{2}$O$_{4}$ and NiFe$_{2}$O$_{4}$ environments at the
interface, consistent with a mixed (Fe,Mn,Ni)$_{3}$O$_{4}$ spinel.
The Ni XAS and XMCD spectra is characteristic of NiFe$_{2}$O$_{4}$,
while the Mn XAS and XMCD spectra is characteristic of
MnFe$_2$O$_4$\cite{Pattrick02}. Additionally, the alignment of the
Ni and Mn moments with respect to the Fe moments in the XMCD is
consistent with MnFe$_{2}$O$_{4}$ and NiFe$_{2}$O$_{4}$. As seen in
Fig 2, the maximum Ni dichroism is parallel to the third peak of the
Fe dichroism, as in bulk NiFe$_{2}$O$_{4}$, and the maximum Mn
dichroism is antiparallel to the third peak of the Fe dichroism, as
in bulk MnFe$_{2}$O$_{4}$\cite{Antonov03}. Furthermore, the lack of
a change in the Mn and Ni XMCD spectra below the NMO T$_C$ may
result from probing the magnetism in an interdiffused region, which
would form the bulk of the XMCD probing depth and, due to the
comparatively high saturation magnetization values of
NiFe$_{2}$O$_{4}$ and MnFe$_{2}$O$_{4}$ with respect to NMO,
overwhelm the NMO dichroism.

In conclusion, isostructural spinel interfaces of Fe$_{3}$O$_{4}$
and NiMn$_{2}$O$_{4}$ exhibit strong interfacial magnetic coupling,
although the NMO T$_{C}$ of 60K is preserved. Element and interface
specific magnetic analysis suggests that this behavior is due to
limited interdiffusion of the Fe, Mn and Ni cations at the
interface, thus creating a spinel solid solution of
(Fe,Mn,Ni)$_{3}$O$_{4}$ that exhibits the magnetic properties of
NiFe$_{2}$O$_{4}$ and MnFe$_{2}$O$_{4}$. Above the NMO T$_{C}$, this
interdiffused region couples to the magnetic Fe$_{3}$O$_{4}$ layers;
however, with the onset of ferrimagnetism in the NMO film, the
interfacial region first becomes frustrated, and then couples to the
magnetically hard NMO. This work is relevant for understanding
magnetic interfacial interactions in spinel-spinel heterostructures.

This work was supported in full by the Director, Office of Science,
Office of Basic Energy Sciences, Division of Materials Sciences and
Engineering, of the U.S. Department of Energy under Contract No.
DE-AC02-05CH11231.

\end{document}